\documentclass[aps,prl,twocolumn,superscriptaddress]{revtex4}
\usepackage{graphicx}
\usepackage{amssymb}
\usepackage{dcolumn}
\usepackage{bm}
\begin{document}
\title{Thickness dependence of the mobility at the LaAlO$_3$ / SrTiO$_3$ interface}
\author{C. Bell}
\affiliation{Department of Advanced Materials Science, University of Tokyo, Kashiwa, Chiba 277-8651, Japan}
\affiliation{Japan Science and Technology Agency, Kawaguchi, 332-0012, Japan}
\author{S. Harashima}
\affiliation{Department of Advanced Materials Science, University of Tokyo, Kashiwa, Chiba 277-8651, Japan}
\author{Y. Hikita}
\affiliation{Department of Advanced Materials Science, University of Tokyo, Kashiwa, Chiba 277-8651, Japan}
\author{H. Y. Hwang}
\affiliation{Department of Advanced Materials Science, University of Tokyo, Kashiwa, Chiba 277-8651, Japan}
\affiliation{Japan Science and Technology Agency, Kawaguchi, 332-0012, Japan}
\date{\today}
\newcommand{\sto}{SrTiO$_3$}          
\newcommand{\stos}{SrTiO$_3$ }          
\newcommand{\stod}{SrTiO$_{3-\delta}$}
\newcommand{\stods}{SrTiO$_{3-\delta}$ }
\newcommand{\lao}{LaAlO$_3$}          
\newcommand{\laos}{LaAlO$_3$ }          
\newcommand{\etal}{{\it et al.}}      %
\newcommand{\ie}{{\it i.e.}}          %
\newcommand{\eg}{{\it e.g.}}          %
\begin{abstract}The electronic transport properties of a series of LaAlO$_3$ / SrTiO$_3$ interfaces were investigated, and a systematic thickness dependence of the sheet resistance and magnetoresistance was found for constant growth conditions. This trend occurs above the critical thickness of four unit cells, below which the LaAlO$_3$ / SrTiO$_3$ interface is not conducting. A dramatic decrease in mobility of the electron gas of nearly two orders of magnitude was observed with increasing LaAlO$_3$ thickness from five to 25 unit cells.\end{abstract}
\maketitle

There is an on-going debate concerning the origin of the conducting layer formed between the two insulators \laos and \stos\cite{ohtomo_nature2004}. Electron doping due to the polar discontinuity between the $\{100\}$ interfaces of \laos and \stos\cite{nakagawa_natmat2006}, atomic diffusion \cite{willmott_prl2007} or the formation of conducting \stods \cite{kalabukhov_prb2007,siemons_prl2007,herranz_prl2007} are all competing possible mechanisms. A central difficulty in achieving consensus about the electronic properties of this system is the variation in growth parameters used by various groups, leading to a wide range of reported properties. In order to both understand the mechanism of the conducting interface, and to control its electronic parameters for possible applications, a thorough knowledge of the growth control parameters and the electronic phase diagram is essential.  

In common with the growth of other thin films by pulsed laser ablation, the substrate temperature and laser energy density are highly influential over the film properties. The importance of oxygen partial pressure during the growth of the \laos layer has also been emphasized \cite{ohtomo_nature2004,brinkman_natmat2007}: a progressive change from metallic to almost insulating behavior at low temperatures was found systematically with increasing with oxygen pressure. Similar metallic samples show superconductivity below 0.3 K \cite{reyren_science2007,caviglia_nature2008}, and some form of magnetic ordering has been suggested as the ground state of the samples showing an insulating tendency \cite{brinkman_natmat2007}, although the exact nature of the electronic ordering is currently unclear. 

The \laos thickness was also shown to be another crucial control parameter \cite{thiel_science2006}. In this case the \laos and \stos interface resistance was $>$1 G$\Omega$ at 300 K for a thickness $d\le3$ unit cells (uc), but became conducting for $d\ge4$ uc. This result can be interpreted to support to the polar discontinuity picture in favour of other doping mechanisms. In that work \cite{thiel_science2006} the 4 uc and 6 uc sample showed a metallic resistance versus temperature, R(T), but there was no report of the properties of the thicker samples up to 15 uc, except for their conductivities and sheet carrier densities at 300 K. In this Letter we report low temperature magnetotransport measurements showing that the metallicity is strongly thickness dependent far above the critical 4 uc value, and that thickness can play an analogous role to oxygen pressure in previous studies.    

Our samples were grown by pulsed laser deposition using a KrF laser in an oxygen pressure of 1.33 mPa, at a repetition rate of 2 Hz. The total laser energy was 20 mJ, and the laser was imaged to a rectangular spot of area 2.3 mm $\times$ 1.3 mm on the single crystal \laos target using an afocal zoom stage.  Each sample was grown on a 5 mm $\times$ 5 mm \stos (100) substrate with TiO$_2$ terminated surfaces \cite{kawasaki_science1994,koster_apl1998}. All substrates in this work were cut from a single 15 mm $\times$ 15 mm \stos substrate to ensure comparable initial surfaces. Before growth the substrates were preannealed at 1223 K for 30 minutes in an oxygen environment of 0.67 mPa. Following this anneal, the substrate temperature was reduced to 1073 K, as measured by an external optical pyrometer. The five \laos thicknesses were 2, 5, 10, 15 and 25 uc, as monitored using {\it in-situ} reflection high-energy electron diffraction (RHEED). Clear RHEED intensity oscillations were observed for all samples, giving a growth rate of $\sim 34\pm2$ laser pulses per uc. \begin{figure}[h]\includegraphics[width=8cm]{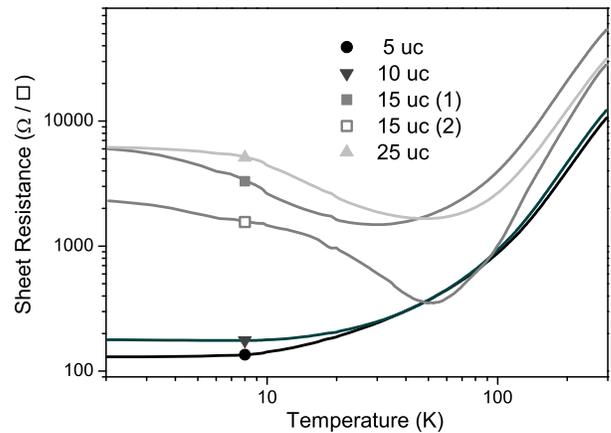}\caption{\label{RTfigure}Resistance per square versus temperature for \laos thicknesses of 5, 10, 15 and 25 uc. Two curves are shown for different parts of the same sample with 15 uc thickness.}\end{figure}

After growth the sample was cooled to room temperature with the chamber filled by $4\times10^4$ Pa of oxygen, with a one hour pause at 873 K. This post annealing step is the same as used elsewhere \cite{thiel_science2006}. The samples were electrically contacted using an ultrasonic wirebonder with aluminium wire, in a Hall bar configuration with voltage contacts $\sim$ 1 mm apart. All measurements were made in a Quantum Design PPMS with the sample normal parallel to the applied magnetic field.  

In Figure \ref{RTfigure} we show the resistance per square versus temperature of the various samples. The 2 uc sample, below the critical thickness, was highly resistive, and its properties will not be discussed further in this Letter. The thickness dependence of the other samples is clear: the thicker samples show an upturn in the resistance at low temperatures, analogous to the high oxygen pressure samples of other groups, whereas the 5 uc and 10 uc sample show metallic behavior down to 2 K. While the other samples showed reproducible data, the 15 uc sample showed some inhomogeneity in the transport for different areas on the same sample, hence we show two R(T) curves to represent this. For clarity in the following discussions, we focus on the sample 15 uc (1), and do not show the magnetotransport data for the sample labeled 15 uc (2). \begin{figure}[h]\includegraphics[width=8cm]{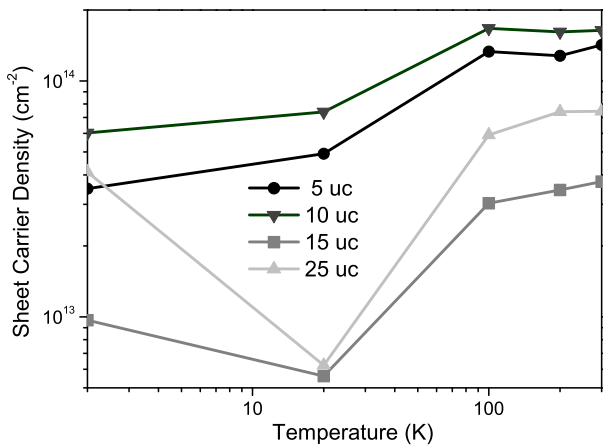}\caption{\label{Hallfigure}Sheet carrier density versus temperature for the four conducting samples. The Hall coefficient was calculated at 1 T.}\end{figure} 

The sheet carrier density versus temperature is shown in Figure \ref{Hallfigure}. This was calculated at 1 T from antisymmetrized Hall measurements. Similar to recent literature the thicker samples show densities close to $1\times10^{13}$ cm$^{-2}$, whereas the more metallic samples approach $1\times10^{14}$ cm$^{-2}$. In this case, there is no clear systematic change with the thickness of the film, although a difference between the two thinner and two thicker samples is apparent. The distinct drop in carrier density around 20 K was also observed elsewhere \cite{suppinfobrinkman} for more resistive samples. Note that since we measure only the sheet carrier density, the thickness of the electron gas, and thus the volume carrier density is unknown and may change systematically with thickness. 

\begin{figure}[h]\includegraphics[width=8cm]{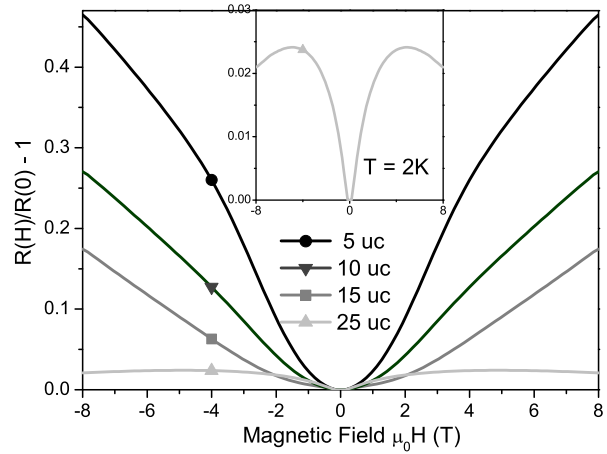}\caption{\label{MRfigure}Symmetrized magnetoresistance at 2 K with magnetic field applied parallel to the substrate normal for the four thicknesses. Inset: Enlarged data for the 25 uc sample, showing clear local maxima.}\end{figure}

Similar to the sheet resistance data, the magnetoresistance at 2 K shows a clear systematic change: a steady reduction from a strong positive magnetoresistance for the thinnest conducting sample, to the curve for 25 uc, which shows a decreasing magnetoresistance at high fields, similar to the negative magnetoresistance observed previously by other groups, (see Inset of Figure \ref{MRfigure}). Again the analogy between oxygen pressure in the latter work, and thickness in the current work is clear. 

\begin{figure}[h]\includegraphics[width=8cm]{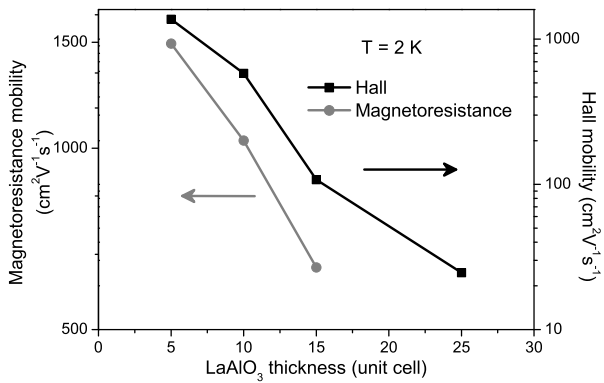}\caption{\label{mobilityfigure}Hall and Magnetoresistance Mobility versus \laos thickness for the four samples at T = 2 K. }\end{figure}

Finally we calculate the electron mobility $\mu$ from both the Hall effect measurement and the magnetoresistance data using the two well known relationships $\mu_H = 1 / eRn$ and $R(B)-R(0) =1+ \zeta (\mu_{mr} B)^2$, where $e$ is the electronic charge, $n$ the sheet carrier density, $B$ the flux density and $\zeta$ is of order unity \cite{schroderbook}. We extracted the magnetoresistance mobility from a low field fit to the $R(B)$ data. Due to the unconventional shape of the $R(B)$ for the 25 uc sample, as mentioned above, we are unable to extract a physically meaningful $\mu_{mr}$ for this sample. These two measures of mobility are plotted in Figure \ref{mobilityfigure}. A clear and striking drop in the Hall mobility $\mu_H$ is apparent as the \laos thickness increases. We also see a clear decrease in the magnetoresistance mobility $\mu_{mr}$, although the rate of change with thickness for $\mu_{mr}$ is lower than for $\mu_H$. Thus while $\zeta = \mu_{mr}/\mu_H$ is close to unity for the 5 uc sample, the 10 and 15 uc samples show an increasing disagreement between $\mu_{mr}$ and $\mu_H$. At present the reason for this difference in behavior of the two measures of mobility is not understood, but an additional scattering mechanism in the electron gas, possibly magnetic in origin, may affect the magnetoresistive measurements and explain this discrepancy.    

Atomic force microscopy over a 20 $\mu$m $\times$ 20 $\mu$m area on the surface of the samples showed clear unit cell steps, but no features could be identified that clearly changed with the thickness of the film. We have observed other thicker films ($>$30 uc), which showed some surface features that suggest cracking associated with the relaxation of the strained \lao, but this was not evident in these samples. In addition to microscopic defects that might be found at the interface \cite{maurice_epl2008}, many theoretical calculations in various systems have suggested a strong link between the transport properties of heterointerfaces such as the \lao/\stos system with the details of the atomic arrangement and bonding lengths in the system (for example \cite{hamann_prb2006,okamoto_prl2006,pentcheva_prb2008}). Such effects may be present in the current experiment.

We note that in the recent literature, the superconducting and gating experiments have tended to use relative thin \laos layers \cite{thiel_science2006, reyren_science2007} (less than 16 uc), whereas those investigating the unusual magnetoresistive properties of the more highly resistive samples are relatively thick (26 uc) \cite{brinkman_natmat2007}. In the former case highly resistive thinner films could not be grown \cite{caviglia_comment}. This is consistent with the data presented in this Letter. However we also note that in different growth conditions, thicker films can be metallic (and superconducting), therefore we cannot assign thickness as the most dominant parameter in this phase space. Nonetheless we clearly can conclude that the \laos thickness, even far above the critical 4 uc value required for conductivity, can have a decisive impact on the mobility and magnetoresistive properties of the electron gas. This thickness dependence can be an additional important parameter for elucidating the microscopic transport mechanisms at the \lao/\stos interface, and assist our understanding and optimization of the electronic properties for future applications. 

We thank A. Caviglia and S. Thiel for useful discussions. CB aknowledges partial funding from the Canon Foundation in Europe.

\end{document}